\documentclass[conference,tbtags,10pt]{IEEEtran}

\usepackage[IEEE]{enderheader}

\newcommand{\Nm}{\v{N}}

\newcommand{\Wm}{\v{W}}
\newcommand{\channel}[1]{\mathsf{#1}}
\newcommand{\Mch}{\channel{M}}

\newcommand{\MWch}{\channel{MW}}
\newcommand{\twon}[1]{2^{n(#1)}}

\newcommand{\twoovern}{\frac{2}{n}}

\newcommand{\CM}[1][sum]{C^{{\scriptscriptstyle (\Mch)}}_{#1}}
\newcommand{\CMW}[1][sum]{C^{{\scriptscriptstyle (\MWch)}}_{#1}}
\newcommand{\CMWs}[1][sum]{\tilde{C}^{{\scriptscriptstyle (\MWch)}}_{#1}}

\begin{document}
\thispagestyle{headings}
\title{The Gaussian Multiple Access Wire-Tap Channel with Collective Secrecy Constraints}
\author{
\begin{minipage}[t]{1.8in}
\centering Ender~Tekin\\\ital{tekin@psu.edu}\\ \vspace{.1in}
\end{minipage}
\begin{minipage}[t]{1.8in}
\centering Aylin~Yener\\\ital{yener@ee.psu.edu}\\ \vspace{.1in}
\end{minipage}\\
{Wireless Communications and Networking Laboratory}\\
{Electrical Engineering Department} \\
{The Pennsylvania State University}  \\
{University Park, PA 16802} \\
\vspace{-.4in}}
\vspace{-.1in}
\maketitle

\begin{abstract}
We consider the Gaussian Multiple Access Wire-Tap Channel (GMAC-WT).  In this scenario,
multiple users communicate with an intended
receiver in the presence of an intelligent and informed
wire-tapper who receives a degraded version of the
signal at the receiver. We define a suitable security measure for this multi-access environment.
We derive an outer bound for the rate region such that secrecy to some pre-determined degree can be maintained. We also find, using Gaussian codebooks, an achievable such secrecy region. Gaussian codewords are shown to achieve the sum capacity outer bound, and the achievable region concides with the outer bound for Gaussian codewords, giving the capacity region when inputs are constrained to be Gaussian.  We present numerical results showing the new rate region and compare it with that of the Gaussian Multiple-Access Channel (GMAC) with no secrecy constraints.
\end{abstract}
\vspace{-.1in}
\section{Introduction}

Shannon, in \cite{shannon:secrecy}, analyzed secrecy systems in communications and he showed that to achieve perfect secrecy of communications, we must
have the conditional probability of the \ital{cryptogram given a message}
independent of the actual transmitted message.

In \cite{wyner:wiretap}, Wyner applied this concept to the
discrete memoryless channel, with a wire-tapper who has access to a degraded version of the intended receiver's signal.  He measured the amount of ``secrecy" using the conditional entropy $\Delta$, the conditional entropy of the transmitted message given the received signal at the wire-tapper.  The region of all possible $(R,\Delta)$ pairs is determined, and the existence of a \ital{secrecy capacity}, $C_s$, for communication below which it is possible to transmit zero information to the
wire-tapper is shown \cite{wyner:wiretap}.

Carleial and Hellman, in \cite{hellman-carleial:wiretap}, showed that
it is possible to send several low-rate messages, each completely
protected from the wire-tapper individually, and use the channel at close to capacity.  The drawback is, in this case, if any of the messages are revealed to the wire-tapper, the others might also be compromised.  In \cite{leung-hellman:gaussianwiretap}, the authors extended Wyner's results to Gaussian channels and also showed that Carleial and Hellman's results in \cite{hellman-carleial:wiretap} also held for the Gaussian channel
\cite{leung-hellman:gaussianwiretap}.
Csisz\'ar and K\"orner, in \cite{csiszar-korner:confbroadcast},
showed that Wyner's results can be extended to weaker, so called ``less noisy" and ``more capable" channels. Furthermore, they analyzed the more general case of sending common information to both the receiver and the wire-tapper, and private information to the receiver only.  More recently, Maurer showed in \cite{maurer:secretkeypublicdiscussion} that a public feedback channel can make secret communications possible even when the secrecy capacity is zero.

In \cite{tekin:ASILOMAR05} we extended these concepts to the GMAC and defined two separate secrecy constraints, which we called \ital{individual} and \ital{collective} secrecy constraints.  We concerned ourselves mainly with the \ital{perfect secrecy rate region} for both sets of constraints.  For the individual constraints, this corresponds to the entropy of the transmitted messages given the received wire-tapper signal and the other users' transmitted signals being equal to the entropy of the transmitted message.  The collective secrecy constraints provided a more relaxed approach and utilized other users' signals as an additional source of secrecy protection.  In this paper, we consider the GMAC-WT and focus on the "collective secrecy constraints" for the GMAC-WT, defined in \cite{tekin:ASILOMAR05} as the normalized entropy of any set of messages conditioned on the wire-tapper's received signal. We consider the general case where a pre-determined level of secrecy is provided. 
Under these constraints, we find an outer bound for the secure rate
region. Using random Gaussian codebooks, we find an achievable
\ital{secure rate region} for each constraint, where users can communicate with
arbitrarily small probability of error with the intended receiver,
while the wire-tapper is kept ignorant to a pre-determined level.  We show that when we constrain ourselves to using Gaussian codebooks, these bounds coincide and give the capacity region for Gaussian codebooks.  Furthermore, it is shown that Gaussian codebooks achieve sum capacity for the GMAC-WT using simultaneous superposition coding, \cite{han-kobayashi:interference}.
We also show that a simple TDMA scheme using the results of \cite{leung-hellman:gaussianwiretap} for the single-user case also achieves sum capacity, but provides a strictly smaller region than shown in this paper.
\vspace{-.07in}

\newcommand{\Csd}[1][\delta]{\ensuremath{\s{C}^{(#1)}}}
\newcommand{\Gsd}[1][\delta]{\ensuremath{\s{G}^{(#1)}}}

\section{System Model and Problem Statement}
\label{sec:system}
\begin{figure}[t]
\centering
\includegraphics[height=1.0in,angle=0]{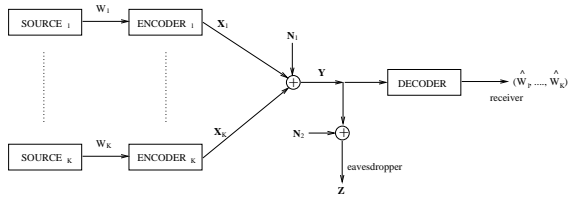}
\caption{The GMAC-WT System Model}
\label{fig:gmacwt}
\end{figure}

We consider $K$ users communicating with a receiver in the presence
of a wire-tapper, as illustrated in Figure \ref{fig:gmacwt}.

Transmitter $j$ chooses a message $W_j$ from a set of equally likely messages
$\{1, \dotsc, M_j\}$. The messages are encoded using $(2^{nR_j},n)$ codes
into $\{X_j^n(W_j)\}$, where $R_j=\ninv \log_2 M_j$. 
The encoded messages are then transmitted,
and the intended receiver and the wire-tapper each get a copy $Y^n$ and
$Z^n$.  We would like to
communicate with the receiver with arbitrarily low probability of
error, while maintaining perfect secrecy, the exact definition of
which will be made precise shortly.

The signal at the intended receiver is given by
\begin{equation}
\Ym = \sum_{j=1}^K \Xm_j + \Nm_1
\end{equation}
and the wire-tapper receives
\begin{equation}
\Zm = \Ym + \Nm_2
\end{equation}
where each component of $\Nm_i \isnormal{0,\nvar_i}, \, i=1,2$.
We also assume the following received power constraints:
\begin{equation}
\ninv \sumton{X_{ji}^2} \le P_{j,max}, \; j=1,\dotsc,K
\end{equation}

\subsection{The Secrecy Measure}
We aim to provide each user with a pre-determined amount of secrecy. To that end, in \cite{tekin:ASILOMAR05}, we used an approach similar to \cite{leung-hellman:gaussianwiretap}, and defined a set of secrecy constraints using the normalized equivocations for sets of users:
\begin{equation}
\Delta_\Ss \triangleq \frac{H(\Wm_\Ss|\Zm)}{H(\Wm_\Ss)} \quad \forall \Ss \subseteq \Ks
\end{equation}
where $\Ks=\{1,\dotsc,K\}$ and $\Wm_\Ss = \{W_j\}_{j \in \Ss}$.

As our secrecy criterion, we require that each user $j \in \{1,\dotsc,K\}$ must satisfy $\Delta_\Ss \ge \delta$ for all sets $\Ss \subseteq \Ks$, and  $\delta \in [0,1]$ is the required level of secrecy.  $\delta=1$ corresponds to \ital{perfect secrecy}, where
the wire-tapper is not allowed to get any information; and $\delta=0$ corresponds to
no secrecy constraint. 
This constraint guarantees that each subset of users maintains a level of secrecy greater than $\delta$.  Since this must be true for all sets of users, collectively the system has at least the same level of secrecy.  However, if a group of users are somehow compromised, the remaining users may also be vulnerable.

\subsection{The $\delta$-secret rate region}
\begin{definition}[Achievable rates with $\delta$-secrecy]
\label{def:achrate}
The rate $K$-tuple $\Rm=(R_1,\dotsc,R_K)$ is said to be achievable with $\delta$-secrecy if for any given $\e>0$ there exists a code of sufficient length \n such that
\begin{align}
\ninv \log_2 M_k &\ge R_k - \e \quad k=1,\dotsc,K\\
\Perr &\le \e \\
\Delta_\Ss &\ge \delta \quad \forall \Ss \subseteq \Ks
\end{align}
where user $k$ chooses one of $M_k$ symbols to transmit according to the uniform distribution, and $\Perr$ is the average probability of error.
We will call the set of all achievable rates with $\delta$-secrecy, the \ital{$\delta$-secret rate region}, and denote it $\Csd$.
\end{definition}

\subsection{Some Preliminary Definitions}
Before we state our results, we define the following quantities for any $\Ss \subseteq \Ks$.
\begin{gather*}
P_\Ss \triangleq \sum_{j \in \Ss} P_j \hspace{1in} R_\Ss \triangleq \sum_{j \in \Ss} R_j \\
\CM[\Ss] \triangleq C\paren{\frac{P_\Ss}{\nvar_1}} \hspace{.7in}  
\CMW[\Ss] \triangleq C\paren{\frac{P_\Ss}{\nvar_1+\nvar_2}} \\
\CMWs[\Ss] \triangleq C\paren{\frac{P_\Ss}{P_{\Ss^c}+\nvar_1+\nvar_2}}
\end{gather*}
where $C(\xi) \triangleq \onehalf \log (1+\xi)$.  The quantities with $\Ss=\Ks$ will sometimes also be used with the subscript \ital{sum}.

\newcommand{\Csdu}{\ensuremath{\hat{\s{C}}^{(\delta)}}}
\newcommand{\Gsdu}{\ensuremath{\hat{\s{G}}^{(\delta)}}}

\section{Outer Bound on the $\delta$-Secret Rate Region}
\label{sec:outer}
In this section, we present an outer bound on the set of achievable $\delta$-secret rates, denoted \Csdu, and explicitly state the outer bound on the achievable sum-rate with $\delta$-secrecy.  We also evaluate this bound assuming we are limited to using Gaussian codebooks for calculation purposes, \Gsdu.

Our main result is presented in the following theorem:
\begin{theorem}
\label{thm:outer}
For the GMAC-WT, the secure rate-tuples $(R_1,\dotsc, R_K)$ such that $\Delta_\Ss \ge \delta$, $\forall \Ss \subseteq \Ks$ must satisfy
\begin{align}
\label{eqn:GMAC}
R_\Ss &\le \CM[\Ss] \\
\label{eqn:outer}
R_\Ss &\le \frac{1}{\delta} \bracket{ \CM[\Ss] - C\paren{\frac{\sum_{j \in \Ss} 2^{\twoovern H(\Xm_j)}}{2 \pi e \paren{P_{\Ss^c}+\nvar_1+\nvar_2}}}}
\end{align}
The set of all $\Rm$ satisfying \eqref{eqn:GMAC} and \eqref{eqn:outer} is denoted \Csdu.
\end{theorem}

\begin{corollary}
\label{cor:outersum}
The sum-rate with $\delta$-secrecy satisfies
\begin{equation}
C_{sum}^{(\delta)} = \sum_{j=1}^K R_j \le \min\braces{\CM, \frac{\CM - \CMW}{\delta}}
\end{equation}
\end{corollary}

\begin{corollary}
\label{cor:outerG}
The rate-tuples with $\delta$-secrecy using Gaussian codebooks must satisfy \eqref{eqn:GMAC} and 
\begin{equation}
\label{eqn:Gdelta}
R_\Ss \le \frac{\CM[\Ss] - \CMWs[\Ss]}{\delta} \quad \forall \Ss \subseteq \Ks
\end{equation}
The set of all such $\Rm$ is denoted \Gsdu.
\end{corollary}

\begin{proof}
See Appendix \ref{app:outerprf}.
\end{proof}
\begin{remark}
Since $\CMW[\Ks]=\CMWs[\Ks]$, Corollary \ref{cor:outerG} indicates that 
Gaussian codebooks have the same upper bound on sum capacity given by Corollary \ref{cor:outersum}.
\end{remark}

\newcommand{\Csdl}{\ensuremath{\check{\s{C}}^{(\delta)}}}
\newcommand{\Gsdl}{\ensuremath{\check{\s{G}}^{(\delta)}}}

\section{Achievable $\delta$-Secret Rate Regions}
\label{sec:achievable}
\subsection{Gaussian Codebooks}
In this section, we find a set of achievable rates using Gaussian codebooks, which we call \Gsdl, and show that Gaussian codebooks achieve the limit on sum capacity.  This region coincides with our previous upper bound evaluated using Gaussian codebooks, \Gsdu, giving the full characterization of the $\delta$-secret rate region using Gaussian codebooks, \Gsd.

\begin{theorem}
\label{thm:ach}
We can transmit with $\delta$-secrecy using Gaussian codebooks at rates satisfying 
\eqref{eqn:GMAC} and \eqref{eqn:Gdelta}.  The region containing all $\Rm$ satisfying these equations is denoted \Gsdl.
\end{theorem}
\begin{corollary}
\label{cor:perfectach}
We can transmit with perfect secrecy ($\delta=1$) using Gaussian codebooks at rates satisfying 
\begin{equation}
R_\Ss \le \CM[\Ss]-\CMWs[\Ss]
\end{equation}
\end{corollary}
\begin{proof} See Appendix \ref{app:achprf}. The corollary was also presented in \cite{tekin:ASILOMAR05}.
\end{proof}
\vspace{-.1in}
\subsection{Time-Division}
We can also use a TDMA scheme and the result of \cite{leung-hellman:gaussianwiretap} to get an achievable region:
\begin{theorem}
Consider this scheme: Let $\alpha_k \in [0,1], \, k=1,\dotsc,K$ and $\sum_{k=1}^K \alpha_k = 1$.  User $k$ only transmits $\alpha_k$ of the time with power $P_{k,max}/\alpha_k$ using the scheme described in \cite{leung-hellman:gaussianwiretap}.  Then, the following set of rates is achievable:
\begin{align}
\label{eqn:RTDMA}
\notag 
\bigcup_{\substack{\v{0} \preceq \vg{\alpha} \preceq \v{1} \\ \sum_{k=1}^K \alpha_k=1}} 
\Big\{\Rm \colon R_k&\le \alpha_k \frac{C\paren{\frac{P_{k,max}}{\alpha_k\nvar_1}}-C\paren{\frac{P_{k,max}}{\alpha_k(\nvar_1+\nvar_2)}}}{\delta} , \\
\vspace{-0.5in}
R_k &\le \alpha_k C \Big( \frac{P_{k,max}}{\alpha_k\nvar_1} \Big), \, k=1,\dotsc,K \Big\}
\vspace{-0.05in}
\end{align}

We will call the set of all $\Rm$ satisfying the above, $\Csd_{TDMA}$.
\end{theorem}
\begin{proof}
Follows directly from \cite[Theorem 1]{leung-hellman:gaussianwiretap}
\end{proof}
\section{Numerical Results and Conclusions}
\label{sec:results}
\newlength{\figsize}
\setlength{\figsize}{1.9in}
\begin{figure}[t]
\centering
\includegraphics[height=\figsize,angle=0]{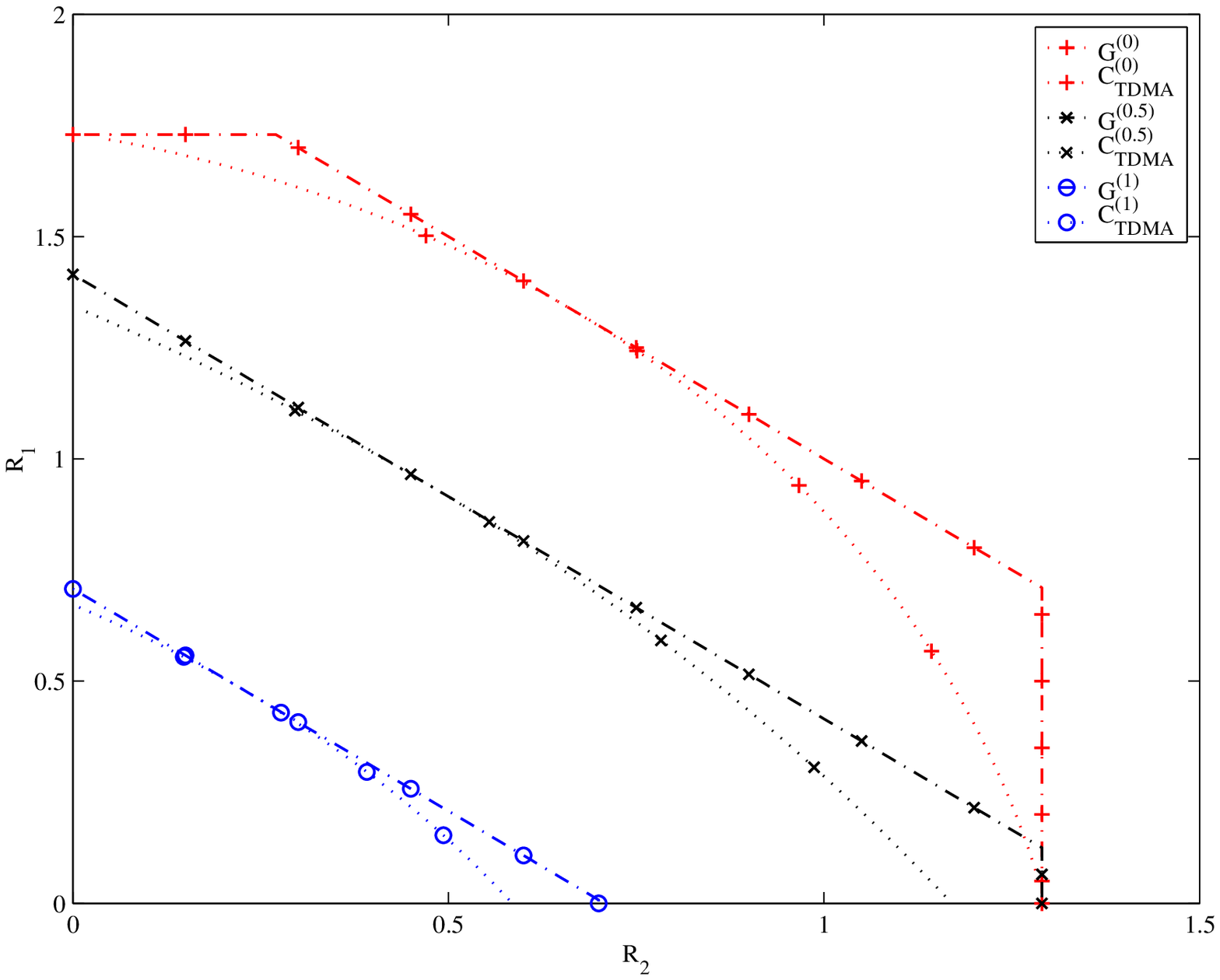}
\vspace{-0.1in}
\caption{Regions for $\delta=0,0.5,1$ and $P_1=10$, $P_2=5$, $\nvar_1=1$, $\nvar_2=2$}
\label{fig:C2}
\vspace{-0.15in}
\end{figure}
\begin{figure}[t]
\centering
\includegraphics[height=\figsize,angle=0]{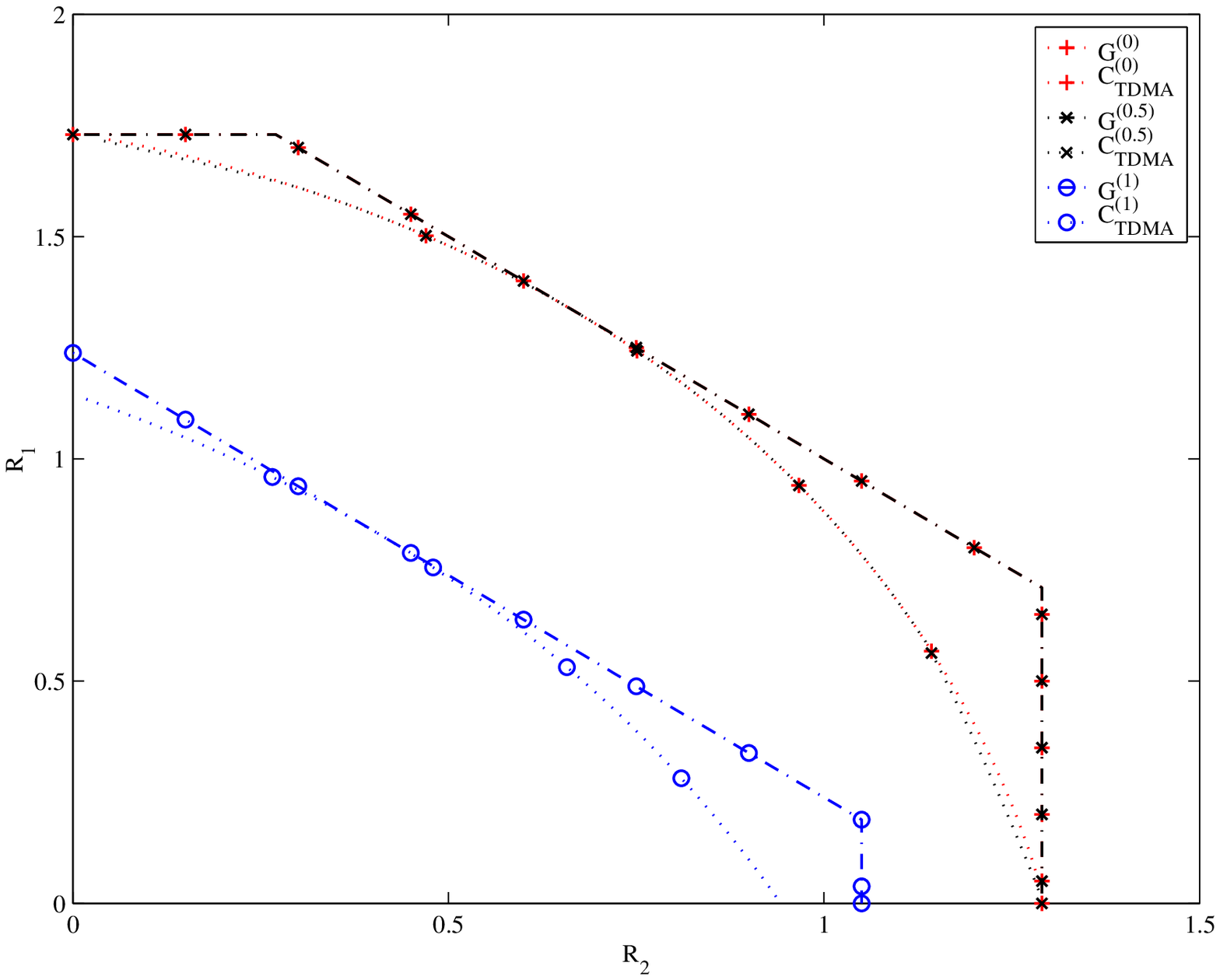}
\vspace{-0.1in}
\caption{Regions for $\delta=0,0.5,1$ and $P_1=10$, $P_2=5$, $\nvar_1=1$, $\nvar_2=7$}
\label{fig:C7}
\vspace{-0.15in}
\end{figure}
\begin{figure}[t]
\centering
\includegraphics[height=\figsize,angle=0]{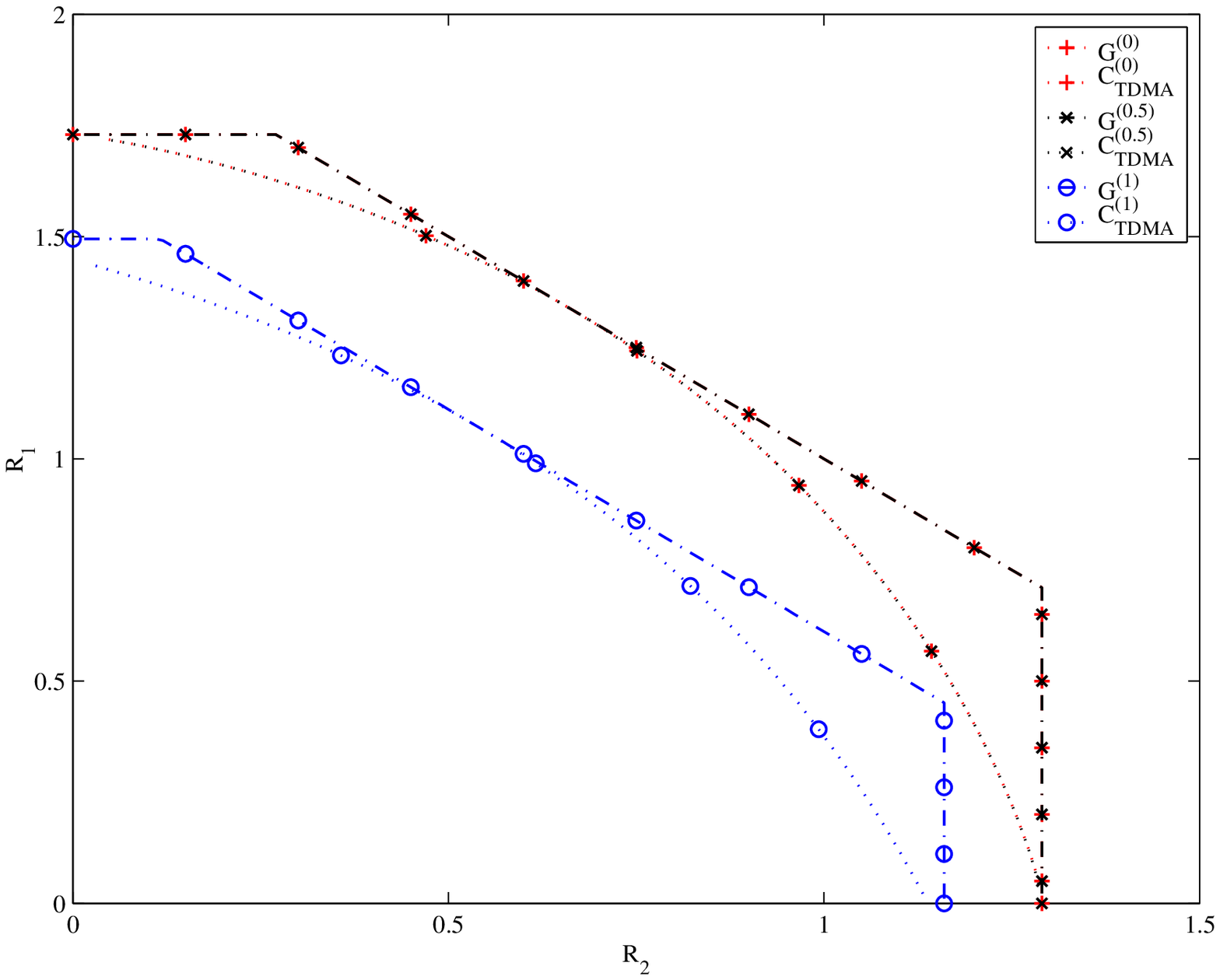}
\vspace{-0.1in}
\caption{Regions for $\delta=0,0.5,1$ and $P_1=10$, $P_2=5$, $\nvar_1=1$, $\nvar_2=20$}
\label{fig:C20}
\vspace{-0.15in}
\end{figure}
\begin{figure}[t]
\centering
\includegraphics[height=\figsize,angle=0]{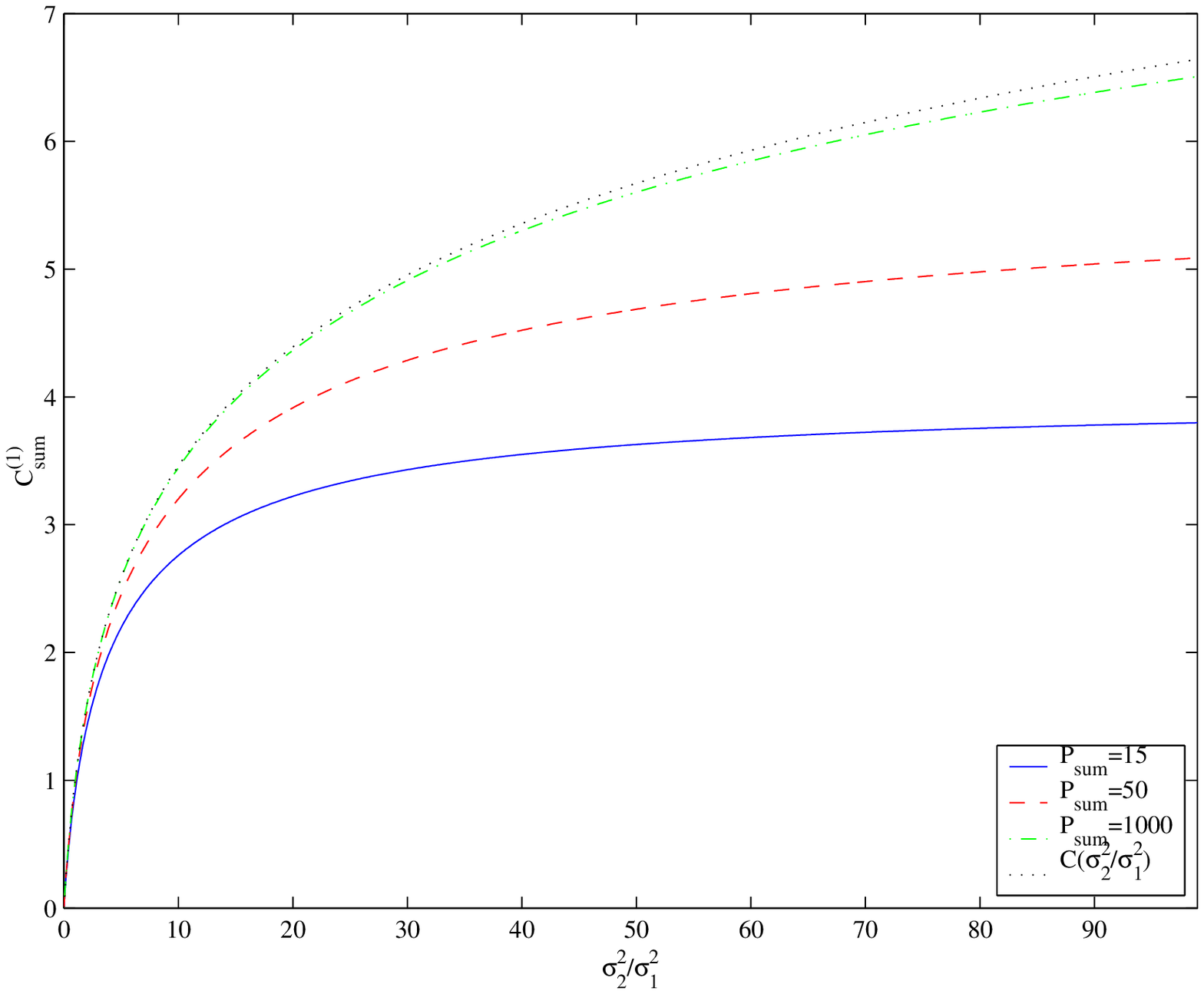}
\vspace{-0.1in}
\caption{$C_{sum}^{(1)}$ vs. $\nvar_2/\nvar_1$.  $C(15)=4,\,C(50)=5.67,\,C(1000)=9.97$}
\label{fig:Cvsn}
\vspace{-0.15in}
\end{figure}

Figures \ref{fig:C2}--\ref{fig:C20} show the shapes of $\Gsd$ for $\delta=0,0.5,1$ for two users.  When $\delta=0$, we are not concerned with secrecy, and the resulting region corresponds to the standard GMAC region, \cite{cover-thomas:IT}.  The region for $\delta=1$ corresponds to the \ital{perfect secrecy} region - transmitting at rates within this region, it is possible to send zero information to the wire-tapper.  The intermediate region, $\delta=0.5$, can be thought of as constraining at least half the transmitted information to be secret. It can be seen that this enlarges the region from the perfect secrecy case.  In Figure \ref{fig:C2}, it is shown that relaxing this constraint may provide a larger region, the limit of which is the GMAC region.  In Figures \ref{fig:C7} and \ref{fig:C20}, however, this region is already equivalent to the GMAC capcity region.  Hence, relaxing our secrecy constraints will not result in further improvement in the set of achievable rates.  Note that it is possible to send at capacity of the GMAC and still provide a non-zero level of secrecy, the minimum value of which depends on how much extra noise the wire-tapper sees.
Also shown in the figures is the regions achievable by the TDMA scheme described in the previous section.  Although TDMA achieves the sum capacity with optimum time-sharing parameters, this region is in general contained within $\Gsd$.

One important point is the dependence of the perfect secrecy region, $\Gsd[1]$, on $\nvar_2$.  It can easily be shown that as $\nvar_2 \toinf$, the perfect secrecy region coincides with the standard GMAC region, $\Gsd[0]$.  Thus, when the wire-tapper sees a much noisier channel than the intended receiver, it is possible to send information with perfect secrecy at close to capacity.  However, when this is not the case, $\Gsd[1]$ is limited by the noise powers regardless of how much we increase the input powers, since  $\limtoinf{P_\Ks} C_{sum}^{(1)} = C(\nvar_2/\nvar_1)$.

Another interesting note is that even when a user does not have any information to send, it can still generate and send random codewords to confuse the eavesdropper and help other users.  This can be seen in Figures \ref{fig:C2} and \ref{fig:C7} as the TDMA region does not end at the ``legs" of $\Gsd$ when $\Gsd$ is not equal to the GMAC capacity region.
\appendices
\allowdisplaybreaks

\section{Outer Bounds}
\label{app:outerprf}
\vspace{-0.05in}
We show that any achievable rate vector, $\Rm$, needs to satisfy Theorem \ref{thm:outer}.  \eqref{eqn:GMAC} is due to the converse of the GMAC coding theorem.  To see \eqref{eqn:outer}, start with a few lemmas:

\begin{lemma}
\label{lem:conv1}
Let $\Xm_\Ss = \{\Xm_k\}_{k \in \Ss}$ where $\Ss \subseteq \Ks$.  Then,
\begin{equation}
R_\Ss \le \frac{1}{n\delta} I(\Xm_\Ss;\Ym|\Zm) +\nu_n \quad \forall \Ss \subseteq \Ks
\end{equation}
where $\nu_n \tozero$ as $\e \tozero$.
\end{lemma}

\begin{proof}
Let $\Ss \subseteq \Ks$ and consider the two inequalities:
\begin{gather}
\label{eqn:lemconv11}
\delta \le \Delta_\Ss 
	=\frac{H(\Wm_\Ss|\Zm)}{\log \paren{\prod_{j \in \Ss} M_j}}
	\le \frac{H(\Wm_\Ss|\Zm)}{n\paren{R_\Ss-\card{\Ss}\e}} \\
\label{eqn:lemconv12}
H(\Wm_\Ss|\Zm,\Ym) \le H(\Wm_\Ss|\Ym) \le H(\Wm_\Ks|\Ym) \le \eta_n
\end{gather}
where \eqref{eqn:lemconv12} follows using Fano's Inequality with $\eta_n \tozero$ as $\e \tozero$ and $n \toinf$. Using \eqref{eqn:lemconv11} and \eqref{eqn:lemconv12}, we can write
\begin{align}
\delta &\le \frac{H(\Wm_\Ss|\Zm)+\eta_n-H(\Wm_\Ss|\Zm,\Ym)}{n\paren{R_\Ss-\card{\Ss}\e}} \\
	&\le \frac{I(\Xm_\Ss;\Ym|\Zm)+\eta_n}{n\paren{R_\Ss-\card{\Ss}\e}}
\end{align}
with the last step using $\Markov{W_\Ss}{\Xm_\Ss}{\Ym}\rightarrow{\Zm}$.  Rearranging  and defining $\nu_n \triangleq \frac{\eta_n}{n\delta}+\card{\Ss}\e$ completes the proof.
\end{proof}

\begin{lemma}[Lemma 10 in \cite{leung-hellman:gaussianwiretap}]
\label{lem:conv3}
Let $\xi = \ninv H(\Ym)$, then,
\begin{equation}
H(\Zm)-H(\Ym) \ge n\phi(\xi) \triangleq \frac{n}{2} \log \bracket{2\pi e \paren{\nvar_2+\frac{1}{2\pi e} 2^{2\xi}}}-n\xi
\end{equation}
\end{lemma}
\vspace{-0.1in}
\begin{corollary}
\label{cor:conv4}
\begin{equation}
H(\Zm) - H(\Ym) \ge \frac{n}{2} \log \paren{1+\frac{\nvar_2}{P_\Ks+\nvar_1}}
\end{equation}
\end{corollary}

\begin{proof}
The lemma is given in \cite{leung-hellman:gaussianwiretap} and its proof is omitted here since it is easily shown using the entropy power inequality, \cite{cover-thomas:IT}.
To see the corollary, write
\begin{equation}
H(\Ym) \le \frac{n}{2} \log \paren{2 \pi e (P_\Ks+\nvar_1)}
\end{equation}
Let $H(\Ym) = n\xi$.  Then, $\xi \le \onehalf \log \paren{2 \pi e (P_\Ks+\nvar_1)}$, and since $\phi(\xi)$ is a non-increasing function of $\xi$, we get $\phi(\xi) \ge 
\phi\paren{\onehalf \log\paren{2\pi e(P_\Ks+\nvar_1)}}$. Then, from \thref{lem:conv3},
\begin{equation}
H(\Zm) - H(\Ym) \ge \frac{n}{2} \log \paren{1+\frac{\nvar_2}{P_\Ks+\nvar_1}}
\vspace{-0.1in}
\end{equation}
\end{proof}
\vspace{-0.1in}
\begin{lemma}
\label{lem:conv2}
For the GMAC-WT,
\begin{equation}
I(\Xm_\Ss;\Ym|\Zm) \le n \CM[\Ss] -n C \paren{\frac{\frac{1}{2 \pi e} \sum_{j \in \Ss} 2^{\twoovern H(\Xm_j)}}	{P_{\Ss^c}+\nvar_1+\nvar_2}}
\end{equation}
\end{lemma}

\begin{corollary}
\label{cor:conv2}
For the GMAC-WT, 
\begin{equation}
I(\Xm_\Ks;\Ym|\Zm) \le n \paren{ \CM - \CMW}
\end{equation}
\end{corollary}
\renewcommand{\ssumton}[1][i]{{\textstyle \sum_{#1=1}^n}}
\renewcommand{\ssum}{{\textstyle \sum}}

\begin{proof} Start by writing
\vspace{-0.05in}
\begin{align}
I(\Xm_\Ss,\Ym|\Zm) \hspace{-0.6in}
	&\hspace{0.6in} =H(\Xm_\Ss|\Zm)-H(\Xm_\Ss|\Ym,\Zm) \\
	&	\label{eqn:lemconv2-1} =H(\Xm_\Ss|\Zm) - H(\Xm_\Ss|\Ym)\\
	&=\bracket{H(\Xm_\Ss)-H(\Xm_\Ss|\Ym)}-\bracket{H(\Xm_\Ss)-H(\Xm_\Ss|\Zm)}\\
	&\notag \le \bracket{H(\Xm_\Ss|\Xm_{\Ss^c})-H(\Xm_\Ss|\Ym,\Xm_{\Ss^c})} \\				
		&\hspace{1in} -\bracket{H(\Xm_\Ss)-H(\Xm_\Ss|\Zm)} \\
	&= I(\Xm_\Ss;\Ym|\Xm_{\Ss^c}) - I(\Xm_\Ss;\Zm) \\
	&= H(\Ym|\Xm_{\Ss^c}) - H(\Ym|\Xm_\Ks)-\bracket{H(\Zm)-H(\Zm|\Xm_\Ss)} \hspace{-0.1in} \\
	&\notag =\ssumton H(Y_i|Y^{i-1},\Xm_{\Ss^c}) 
		- \ssumton H(Y_i|Y^{i-1},\Xm_\Ks) \\
		&\hspace{1in} -\bracket{H(\Zm)-H(\Zm|\Xm_\Ss)} \\
	&\notag \label{eqn:lemconv2-2}\le \ssumton H(Y_i|\Xm_{\Ss^c,i}) 
		- \ssumton H(Y_i|\Xm_{\Ks,i}) \\
		&\hspace{1in} -\bracket{H(\Zm)-H(\Zm|\Xm_\Ss)} \\
	&\notag \le \ssumton \onehalf \log \bracket{2 \pi e \paren{P_\Ss+\nvar_1}} 
		-\ssumton \onehalf \log \paren{2 \pi e \nvar_1} \\
	&\hspace{1in} -\bracket{H(\Zm)-H(\Zm|\Xm_\Ss)} \\
	\label{eqn:lemconv3prf1} &=nC \paren{P_\Ss/\nvar_1} 
		-\bracket{H(\Zm)-H(\Zm|\Xm_\Ss)}
\end{align}
where \eqref{eqn:lemconv2-1} follows from $\Markov{\Xm_\Ss}{\Ym}{\Zm}$ and \eqref{eqn:lemconv2-2} follows using the memoryless property of $\Mch$.  For the term in brackets, start by using the entropy power inequality:
\begin{align}
2^{\twoovern H(\Zm)} &\ge 2^{\twoovern H(\Zm|\Xm_\Ss)} + \ssum_{j \in \Ss} 2^{\twoovern H(\Xm_j)} \\
\label{eqn:lemconv3prf2}
2^{\twoovern H(\Zm) - \twoovern H(\Zm|\Xm_\Ss)} &\ge 1 + 
	2^{-\twoovern H(\Zm|\Xm_\Ss)} \sum_{j \in \Ss} 2^{\twoovern H(\Xm_j)}
\end{align}

\vspace{-0.2in}
Then,
\begin{align}
2^{\twoovern H(\Zm|\Xm_\Ss)} &= 2^{\twoovern \sumton H(Z_i|Z^{i-1},\Xm_\Ss)} \\
	&\le 2^{\twoovern \sumton H(Z_i|\Xm_{\Ss,i})} \\
	&\le 2^{\twoovern \sumton \onehalf \log \paren{2 \pi e (P_{\Ss^c}+\nvar_1+\nvar_2)}}\\
	&=2 \pi e (P_{\Ss^c}+\nvar_1+\nvar_2 )
\end{align}

\vspace{-0.1in}
Using this in \eqref{eqn:lemconv3prf2}, and taking the log we get,
\begin{equation}
\label{eqn:lemconv3prf3}
\hspace{-0.01in} H(\Zm)-H(\Zm|\Xm_\Ss) \ge \frac{n}{2} \log \paren{1 + \frac{\frac{1}{2 \pi e}\sum_{j \in \Ss} 2^{\twoovern H(\Xm_j)}}{P_{\Ss^c}+\nvar_1+\nvar_2}}
\hspace{-0.01in}
\end{equation}
which, with \eqref{eqn:lemconv3prf1} completes the proof.
To see the corollary,
\begin{align}
I(\Xm_\Ks;\Ym|\Zm) \hspace{-.7in}&\hspace{.7in}
	=H(\Xm_\Ks|\Zm) - H(\Xm_\Ks|\Ym,\Zm) \\
	&\label{eqn:lemconv3-1} =H(\Xm_\Ks|\Zm) - H(\Xm_\Ks|\Ym)  \\
	&\notag =\bracket{H(\Zm|\Xm_\Ks) + H(\Xm_\Ks) - H(\Zm)} \\
		&\hspace{1in} -\bracket{H(\Ym|\Xm_\Ks) + H(\Xm_\Ks) - H(\Ym)} \hspace{-1in}\\
	&=\bracket{ H(\Zm|\Xm_\Ks) - H(\Ym|\Xm_\Ks)} - \bracket{H(\Zm)-H(\Ym)} \\
	&\label{eqn:lemconv3-2} =\sumton \bracket{ H(Z_i|\Xm_{\Ks,i}) - H(Y_i|\Xm_{\Ks,i})} - 			\bracket{H(\Zm)-H(\Ym)} \hspace{-.1in}\\
	&=\notag \bracket{ \frac{n}{2} \log \paren{2\pi e (\nvar_1+\nvar_2)} 
		- \frac{n}{2} \log \paren{2\pi e \nvar_1}}\\
		&\hspace{1in} - \bracket{H(\Zm)-H(\Ym)} \\
	&\label{eqn:lemconv3-3}\le \frac{n}{2} \log \paren{1+\frac{\nvar_2}{\nvar_1}} -
		\frac{n}{2} \log \paren{1+\frac{\nvar_2}{P_\Ks+\nvar_1}} \\
	&=\CM - \CMW
\end{align}
where \eqref{eqn:lemconv3-1} is due to $\Markov{\Xm_\Ks}{\Ym}{\Zm}$ and \eqref{eqn:lemconv3-2} to the memory-lessness of the channels.  \eqref{eqn:lemconv3-3} follows from Corollary \ref{cor:conv4}.
\end{proof}

This and Lemma \ref{lem:conv1}, complete the proof of \thref{thm:outer}.

Corollary \ref{cor:outersum} follows from Corollary \ref{cor:conv2} and Lemma \ref{lem:conv1}. 

Corollary \ref{cor:outerG} follows simply with $H(\Xm_j)= \frac{n}{2} \log 2 \pi e P_j$.

\newcommand{\Xc}{\mathfrak{X}}
\newcommand{\Wmi}[1][s]{\Wm^{{\scriptscriptstyle ({\scriptstyle #1})}}}
\newcommand{\XmS}{\Xm_\Sigma}
\newcommand{\Deltai}[1][s]{\Delta^{{\scriptscriptstyle ({\scriptstyle #1})}}}

\vspace{-0.075in}
\section{Achievable Rates}
\label{app:achprf}
\vspace{-0.06in}
Let $\Rm=(R_1,\dotsc,R_K)$ satisfy \eqref{eqn:GMAC} and \eqref{eqn:Gdelta}.
For each user $k \in \Ks$, consider the scheme:
\begin{IEEEenumerate}[
\setlength{\topsep}{0.0in}
\setlength{\parskip}{0in}
\setlength{\labelindent}{0.06in}
\setlength{\labelwidth}{0.05in}
\setlength{\labelsep}{3pt}]
\item Let $M_k=\twon{R_k-\e'}$ where $0 \le \e' < \e$.  Let $M_k=M_{ks}M_{k0}$ where $M_{ks}=M_k^{\mu_k}, \, M_{k0}=M_k^{1-\mu_k}$, and $\mu_k \ge \delta$ will be chosen later. Then, $R_k=R_{ks}+R_{k0}+\e'$ where	$R_{ks}=\ninv \log M_{ks}$ and $R_{k0}=\ninv \log M_{k0}$.  We can choose $\e'$ and $n$ to ensure that $M_{ks},M_{k0}$ are integers.

\item	Generate $3$ codebooks $\Xc_{ks},\Xc_{k0}$ and $\Xc_{kx}$.  $\Xc_{ks}$ consists of $M_{ks}$	codewords, each component of which is drawn $\isnormal{0,\lambda_{ks}P_k -\varepsilon}$. Codebook $\Xc_{k0}$ has $M_{k0}$ codewords with each component randomly drawn $\isnormal{0,\lambda_{k0} P_k-\varepsilon}$ and $\Xc_{kx}$ has $M_{kx}$ codewords with each component randomly drawn $\isnormal{0,\lambda_{kx} P_k-\varepsilon}$ where $\varepsilon$ is an arbitrarily small number to ensure that the power constraints on the codewords are satisfied with high probability and $\lambda_{ks}+\lambda_{k0}+\lambda_{kx}=1$. Define $R_{kx}=\ninv \log M_{kx}$ and $M_{kt}=M_k M_{kx}$.

\item Each message $W_k \in \{1,\dotsc,M_k\}$ is mapped into a message vector	$\Wm_k=(W_{ks},W_{k0})$ where $W_{ks}\in \{1,\dotsc,M_{ks}\}$ and $W_{k0}\in \{1,\dotsc,M_{k0}\}$.	Since $W_k$ is uniformly chosen, $W_{ks},W_{k0}$ are also uniformly distributed.

\item To transmit message $W_k \in \{1,\dotsc,M_k\}$, user $k$ finds the $2$ codewords 
corresponding to components of $\Wm_k$ and also uniformly chooses a codeword from $\Xc_{kx}$. He then adds all these codewords and transmits the resulting codeword, $\Xm_k$, so that we are actually transmitting one of $M_{kt}$ codewords. Let $R_{kt} = \ninv \log M_{kt} +\e'= R_{ks}+R_{k0}+R_{kx}+\e'$. 
\end{IEEEenumerate}

We will choose the rates such that for all $\Ss \subseteq \Ks$,
\begin{align}
\label{eqn:achRs} &\ssum_{k \in \Ss} R_{ks} = \ssum_{k \in \Ss} \mu_k R_k 
	\le \CM[\Ss]-\CMWs[\Ss]\\
\label{eqn:achR0} &\ssum_{k=1}^K \bracket{R_{k0}+R_{kx}} 
	= \ssum_{k=1}^K \bracket{(1-\mu_k)R_k+R_{kx}} = \CMW \qquad \raisetag{.15in}\\
\label{eqn:achRt} &\ssum_{k \in \Ss} R_{kt} 
	= \ssum_{k \in \Ss} \bracket{R_k+R_{kx}} \le \CM[\Ss]
\end{align}
From \eqref{eqn:achRt} and the GMAC coding theorem, with high probability the receiver can decode the codewords with low probability of error. To show $\Delta_\Ss \ge \delta, \; \forall \Ss \subseteq \Ks$, we concern ourselves only with MAC sub-code $\{\Xc_{ks}\}_{k=1}^K$.  From this point of view, the coding scheme described is equivalent to each user $k \in \Ks$ selecting one of $M_{ks}$ messages, and sending a uniformly chosen codeword from among $M_{k0}M_{kx}$ codewords for each.  Let $\Wmi_\Ss = \{W_{ks}\}_{k \in \Ss}$ and $\Deltai_\Ss=\frac{H(\Wmi_\Ss|\Zm)}{H(\Wmi_\Ss)}$ and define $\XmS=\sum_{k=1}^K \Xm_k$. For $\Ks$ write
\vspace{-0.05in}
\begin{align}
\Deltai_{\Ks}
	&=\frac{H(\Wmi_\Ks|\Zm)}{H(\Wmi_\Ks)}=\frac{H(\Wmi_\Ks,\Zm)-H(\Zm)}{H(\Wmi_\Ks)}\\
	&= \frac{H(\Wmi_\Ks,\XmS,\Zm)-H(\XmS|\Wmi_\Ks,\Zm)-H(\Zm)}{H(\Wmi_\Ks)} \\
	&\notag =\frac{H(\Wmi_\Ks)+H(\Zm|\Wmi_\Ks,\XmS)-H(\Zm)}{H(\Wmi_\Ks)} \\
		&\hspace{.5in} +\frac{H(\XmS|\Wmi_\Ks)-H(\XmS|\Wmi_\Ks,\Zm)}{H(\Wmi_\Ks)} \\
	&\label{eqn:achprf1}= 1- \frac{I(\XmS;\Zm)-I(\XmS;\Zm|\Wmi_\Ks)}
		{n \big( \sum_{k=1}^K R_{ks} \big)} 
\end{align}
where we used $\Markov{\Wmi_\Ks}{\XmS}{\Zm} \Rightarrow H(\Zm|\Wmi_\Ks,\XmS)=H(\Zm|\XmS)$ to get \eqref{eqn:achprf1}.
We will consider the two terms individually.  First, we have the trivial bound due to channel capacity: 
\begin{equation}
\label{eqn:achprf3}
I(\XmS;\Zm) \le n\CMW
\vspace{-0.03in}
\end{equation}

\vspace{-0.05in}
$I(\XmS;\Zm|\Wmi_\Ks) = H(\XmS|\Wmi_\Ks)-H(\XmS|\Wmi_\Ks,\Zm)$.
Since user $k$ sends one of $M_{k0}M_{kx}$ codewords for each message,
\vspace{-0.05in}
\begin{align}
H(\XmS|\Wmi_\Ks) &= \log \big({\textstyle \prod}_{k=1}^K M_{k0}M_{kx} \big)\\
	\label{eqn:achprf4a}&= n \ssum_{k=1}^K \bracket{(1-\mu_k)R_k+R_{kx}}
\end{align}

\vspace{-0.05in}
We can also write
\vspace{-0.05in}
\begin{equation}
\label{eqn:achprf4b}
H(\XmS|\Wmi_\Ks,\Zm) \le n\eta'_n
\vspace{-0.1in}
\end{equation}
where $\eta'_n \tozero$ as $n \toinf$ since, with high probability, the eavesdropper can decode $\XmS$ given $\Wmi_\Ks$ due to \eqref{eqn:achR0}.
Using \eqref{eqn:achRs}, \eqref{eqn:achR0}, \eqref{eqn:achprf3}, \eqref{eqn:achprf4a} and \eqref{eqn:achprf4b} in \eqref{eqn:achprf1}, we get
\vspace{-0.05in}
\begin{align}
\hspace{-0.05in} \Deltai_\Ks 
	&\ge 1-\frac{\CMW-\sum_{k=1}^K\bracket{(1-\mu_k)R_k+R_{kx}}+\eta'_n}{\CM-\CMW}\\ 
	\label{eqn:G1}&= 1-\frac{\eta'_n}{\CM-\CMW}\rightarrow 1\;\text{as }\eta'_n \tozero
\vspace{-0.03in}
\end{align}
Then,
\vspace{-0.05in}
\begin{align}
H(\Wmi_\Ks|\Zm) &= H(\Wmi_\Ks) \\
H(\Wmi_\Ss|\Zm) + H(\Wmi_{\Ss^c}|\Zm) &\ge H(\Wmi_\Ss)+H(\Wmi_{\Ss^c})
\end{align}

\vspace{-0.05in}
As conditioning reduces entropy, we have $H(\Wmi_\Ss|\Zm) \le H(\Wmi_\Ss)$ and 
$H(\Wmi_{\Ss^c}|\Zm) \le H(\Wmi_{\Ss^c})$.  Then, from the above equation we conclude that we must have $H(\Wmi_\Ss)=H(\Wmi_\Ss|\Zm), \; \forall \Ss \subset \Ks$.  This makes $\Deltai_\Ss=1 \; \forall \Ss \subset \Ks$.  The proof is completed by noting that
\vspace{-0.05in}
\begin{equation}
\label{eqn:achprf0}
\Delta_\Ss \ge \frac{H(\Wmi_\Ss|\Zm)}{H(\Wm_\Ss)} 
	= \frac{H(\Wmi_\Ss)}{H(\Wm_\Ss)} 
	= \frac{\ssum_{k \in \Ss} \mu_k R_k}{\ssum_{k \in \Ss} R_k}
	\ge \delta
	\vspace{-0.05in}
\end{equation}

We can think of $\{W_{ks}\}$ as the ``protected" messages and $\{W_{k0}\}$ as the ``unprotected" messages.
The corollary is apparent from \eqref{eqn:G1}, and also follows as \eqref{eqn:Gdelta} implies \eqref{eqn:GMAC} if $\delta=1$.
\vspace{-0.13in}
\vspace{-0.1in}
\bibliographystyle{IEEEtran}
\bibliography{IEEEabrv_mod,etekin_full}
\end{document}